\documentclass[prc,twocolumn]{revtex4}
\usepackage{amssymb}

\usepackage{amsmath}
\usepackage{graphicx}

\begin{document}

\preprint{}
\title{Constraints on the $^{22}$Ne($\alpha $,\textit{n})$^{25}$Mg \textit{s}%
-process neutron source from analysis of $^{nat}$Mg+\textit{n} total and $%
^{25}$Mg(\textit{n},$\gamma $) cross sections }
\author{P. E. Koehler}
\email{koehlerpe@ornl.gov}
\affiliation{Physics Division, Oak Ridge National Laboratory, Oak Ridge, Tennessee 37831}
\date{\today}

\begin{abstract}
The $^{22}$Ne($\alpha $,\textit{n})$^{25}$Mg reaction is thought to be the
neutron source during the \textit{s} process in massive and intermediate
mass stars as well as a secondary neutron source during the \textit{s}
process in low mass stars. Therefore, an accurate determination of this rate
is important for a better understanding of the origin of nuclides heavier
than iron as well as for improving \textit{s}-process models. Also, the 
\textit{s} process produces seed nuclides for a later \textit{p} process in
massive stars, so an accurate value for this rate is important for a better
understanding of the \textit{p} process. Because the lowest observed
resonance in direct $^{22}$Ne($\alpha $,\textit{n})$^{25}$Mg measurements is
considerably above the most important energy range for \textit{s}-process
temperatures, the uncertainty in this rate is dominated by the poorly known
properties of states in $^{26}$Mg between this resonance and threshold.
Neutron measurements can observe these states with much better sensitivity
and determine their parameters (except $\Gamma _{\alpha }$) much more
accurately than direct $^{22}$Ne($\alpha $,\textit{n})$^{25}$Mg
measurements. I have analyzed previously reported $^{\text{nat}}$Mg+\textit{n%
} total and $^{25}$Mg(\textit{n},$\gamma $) cross sections to obtain a much
improved set of resonance parameters \ for states in $^{26}$Mg between
threshold and the lowest observed $^{22}$Ne($\alpha $,\textit{n})$^{25}$Mg
resonance, and an improved estimate of the uncertainty in the $^{22}$Ne($%
\alpha $,\textit{n})$^{25}$Mg reaction rate. For example, definitely two,
and very likely at least four, of the states in this region have natural
parity and hence can contribute to the $^{22}$Ne($\alpha $,\textit{n})$^{25}$%
Mg reaction, but two others definitely have non-natural parity and so can be
eliminated from consideration. As a result, a recent evaluation in which it
was assumed that only one of these states has natural parity has
underestimated the reaction rate uncertainty by at least a factor of ten,
whereas evaluations that assumed all these states could contribute probably
have overestimated the uncertainty.
\end{abstract}

\pacs{}
\maketitle

\section{Introduction}

During helium-burning and, perhaps, carbon-burning phases in massive and
intermediate mass stars, the $^{22}$Ne($\alpha $,\textit{n})$^{25}$Mg
reaction is thought to be the neutron source driving the synthesis of
nuclides in the A$\approx $60-90 mass range during the slow-neutron-capture (%
\textit{s}) process \cite{Ka99b,Ho2001}. The \textit{s} process in these
stars also can modify the abundances of several lighter nuclides. The $^{22}$%
Ne($\alpha $,\textit{n})$^{25}$Mg reaction also acts as a secondary neutron
source during the \textit{s} process in low-mass asymptotic giant branch
(AGB) stars during which roughly half the abundances of nuclides in the A$%
\approx $90 - 209 range are thought to be synthesized \cite{St95}. Although
the overall neutron exposure due to this reaction in AGB stars is much
smaller than that due to $^{13}$C($\alpha $,\textit{n})$^{16}$O, the neutron
density as well as the temperature are much higher during the $^{22}$Ne($%
\alpha $,\textit{n})$^{25}$Mg phase, resulting in important modifications to
the final \textit{s}-process abundances. Massive stars during their later
burning stages are also the leading candidates for the production of the
rare neutron-deficient isotopes of nuclides in the A$\gtrsim $90 mass range
through the so-called \textit{p} process. Because the \textit{s} process in
these stars produces seed nuclides for a later \textit{p} process, the size
of the $^{22}$Ne($\alpha $,\textit{n})$^{25}$Mg reaction rate used in the
stellar model can have a significant effect on the predicted abundances of
the \textit{p} isotopes \cite{Co2000}.

There have been several attempts to determine the rate for this reaction
either through direct $^{22}$Ne($\alpha $,\textit{n})$^{25}$Mg measurements %
\cite{Dr91,Ha91,Dr93,Gi93,Ja2001} or indirectly via $^{26}$Mg($\gamma $,%
\textit{n})$^{25}$Mg \cite{Be69} or charged-particle transfer reactions \cite%
{Gi93}. However, direct measurements have suffered from relatively poor
resolution as well as the fact that the cross section is extremely small at
the lower energies corresponding to \textit{s}-process temperatures.
Indirect methods have also suffered from limited sensitivity and relatively
poor resolution. This rate, in principle, could be determined via the
inverse $^{25}$Mg(\textit{n},$\alpha $)$^{22}$Ne reaction, but the small
size of the cross section in the relevant energy range makes these
measurements exceedingly difficult, so no results have been reported. As a
result, various evaluations \cite{Dr93,Ca88,Ka94,An99} of this rate show
considerable differences and all but the most recent \cite{Ja2001} recommend
uncertainties much larger than needed to adequately constrain astrophysical
models. Because the lowest observed resonance (\textit{E}$_{\alpha }$=832
keV, which corresponds to \textit{E}$_{n}$=235 keV in the inverse reaction)
in direct $^{22}$Ne($\alpha $,\textit{n})$^{25}$Mg measurements is
considerably above the most important energy range for \textit{s}-process
temperatures, the uncertainty in this rate is dominated by the poorly known
properties of states in $^{26}$Mg between this resonance and threshold.
Because both $^{22}$Ne and $^{4}$He have $J^{\pi }$=0$^{+}$, only natural
parity (0$^{+}$, 1$^{-}$, 2$^{+}$..) states in $^{26}$Mg can participate in
the $^{22}$Ne($\alpha $,\textit{n})$^{25}$Mg reaction, so only a subset of $%
^{26}$Mg states in the relevant energy range observed via neutron reactions
can contribute to the reaction rate. Most evaluations have made the
assumption that either all, or only one, of the known states \cite{Mu81,En98}
in this region can contribute to the reaction rate. For example, in the
recent NACRE \cite{An99} evaluation, all known states in this region were
considered when calculating the uncertainty whereas in the most recent
report \cite{Ja2001} only one state was assumed to have natural parity. As a
result, the recommended uncertainties in the NACRE evaluation are much
larger than those in Ref. \cite{Ja2001}. For example, at $T$=0.2 GK, the
NACRE uncertainty is approximately 300 times larger than that given in Ref. %
\cite{Ja2001}.

Much of the information about states in $^{26}$Mg in the relevant energy
region comes from neutron measurements \cite{Ne59,Si74,We76}. In principle,
the combination of neutron total and capture cross section measurements on $%
^{25}$Mg can determine all of the relevant resonance parameters (\textit{E}$%
_{r}$, $J^{\pi }$, $\Gamma _{n}$, and $\Gamma _{^{_{\gamma }}}$) except $%
\Gamma _{\alpha }$ with much better sensitivity and to greater precision
than other techniques. Both high-resolution $^{nat}$Mg+\textit{n} total and $%
^{25}$Mg(\textit{n},$\gamma $)$^{26}$Mg cross sections have been reported %
\cite{We76} and some resonance parameters were extracted from these data.
However, the resonance analysis was rather limited and it is possible to
extract much more information using current techniques. For example,
resonance shapes and peak heights in the total cross section should allow
the extraction of \textit{J}$^{\pi }$ values, but no definite assignments
were made in Ref. \cite{We76} for states in $^{26}$Mg. Also, it should be
possible to determine the partial widths for many of the resonances, but
only five $\Gamma _{n}$ and no $\Gamma _{\gamma }$ values were reported for
the 17 $^{25}$Mg+\textit{n} resonances reported in Ref. \cite{We76}. Partial
width information can be particularly valuable in assessing the relative
strengths of the competing $^{22}$Ne($\alpha $,\textit{n})$^{25}$Mg and $%
^{22}$Ne($\alpha $,$\gamma $)$^{26}$Mg reactions in stars, and hence the
efficiency of the \textit{s}-process neutron source.

I have analyzed previously reported \cite{We76} $^{\text{nat}}$Mg+\textit{n}
total and $^{25}$Mg(\textit{n},$\gamma $)$^{26}$Mg cross sections using the
multi-level, multi-channel $\mathcal{R}$-matrix code SAMMY \cite{La2000} to
obtain a much improved set of resonance parameters \ for states from
threshold through the lowest observed $^{22}$Ne($\alpha $,\textit{n})$^{25}$%
Mg resonances. In the next section, I describe the data and analysis
technique used. In section \ref{CompSec}, I compare the new results to
previous analyses of neutron data as well as resonance parameter information
from $^{22}$Ne($\alpha $,\textit{n})$^{25}$Mg, $^{26}$Mg($\gamma $,\textit{n}%
)$^{25}$Mg, and $^{22}$Ne($^{6}$Li,\textit{d})$^{26}$Mg measurements. In
section \ref{ReacRate}, I use the new resonance parameters together with
recently reported \cite{Ja2001} upper limits for the $^{22}$Ne($\alpha $,%
\textit{n})$^{25}$Mg cross section to compute the uncertainty in this rate
at \textit{s}-process temperatures. I conclude that the most recent report %
\cite{Ja2001} of this reaction rate has underestimated the uncertainty by at
least a factor of 10 and that high resolution $^{25}$Mg+\textit{n} total
cross section measurements would be invaluable in further refining the
uncertainty in this important reaction rate.

\section{Data and $\mathcal{R}$-matrix Analysis}

The best data for the present purposes are those of Ref. \cite{We76}. The
data consist of very high resolution $^{nat}$Mg+\textit{n} total and
high-resolution $^{25}$Mg(\textit{n},$\gamma $)$^{26}$Mg cross sections
measured using time-of-flight techniques at the white neutron source of the
Oak Ridge Electron Linear Accelerator (ORELA) facility \cite{Pe82,Bo90,Gu97b}%
. Total cross sections were measured using a relatively thick (0.2192 at/b)
metallic Mg sample and a plastic scintillator detector on a 200-m flight
path. The neutron capture measurements were made using a thin (0.030 at/b),
97.87\% enriched sample on a 40-m flight path, and employed the pulse-height
weighting technique using fluorocarbon scintillators to detect the $\gamma $
rays. Although total cross section measurements have been reported \cite%
{Ne59} using an enriched $^{25}$Mg sample, the data are of much lower
resolution and precision than those of Ref. \cite{We76}.

The original, unaveraged $^{nat}$Mg+\textit{n} transmission data were
obtained \cite{Ha2001} to preserve the best resolution so that the best
possible parameters could be obtained from fitting the data. The data
between resonances were averaged to speed up the fitting process. Only a
subset of the $^{25}$Mg(\textit{n},$\gamma $)$^{26}$Mg data, corresponding
to energy regions near the resonances reported in Ref. \cite{We76} and
extending to only \textit{E}$_{n}$=275 keV, could be located \cite{Ma2001}.
These data were corrected by a factor of 0.9325 as recommended in Ref. \cite%
{Ma81}. The $^{24,26}$Mg(\textit{n},$\gamma $)$^{25,27}$Mg data of Ref. \cite%
{We76} could not be found.

The data were fitted with the $\mathcal{R}$-matrix code SAMMY \cite{La2000}
to extract resonance parameters. All three stable Mg isotopes were included
in the analysis because the sample for the total cross section measurements
was natural Mg (78.99\% $^{24}$Mg, 10.00\% $^{25}$Mg, and 11.01\% $^{26}$%
Mg). Orbital angular momenta up to and including \textit{d}-waves were
considered. Radii of 4.9 fm were used in all $^{25}$Mg+\textit{n} channels
as well as the $^{24}$Mg+\textit{n} \textit{d}-wave channels, and a radius
of 4.3 fm was used in all $^{26}$Mg+\textit{n} channels. Because $^{24}$Mg
is the major isotope in natural Mg and because the \textit{s}- and \textit{p}%
-wave penetrabilities are considerably larger than \textit{d}-wave at these
energies, \ the \textit{s}- and \textit{p}-wave radii for $^{24}$Mg+\textit{n%
} were allowed to vary while fitting the total cross sections. The fitted
radii were 5.88 fm and 4.17 fm for \textit{s}- and \textit{p}-waves,
respectively, in $^{24}$Mg+\textit{n}. All resonances up to the highest
energy given in Ref. \cite{We76} (1.754 MeV) were included in the $\mathcal{R%
}$ matrix although I did not attempt to fit the total cross section data
above 500 keV because the unavailability of $^{25}$Mg(\textit{n},$\gamma $)$%
^{26}$Mg data above 275 keV made it increasingly difficult to assign $^{25}$%
Mg+\textit{n} resonances at the higher energies. The parameters of Ref. \cite%
{We76}, supplemented by those in Ref. \cite{Mu81} in some cases, were used
as starting points in the fitting process.

\begin{figure}
\includegraphics*[height=85mm,width=85mm,keepaspectratio] {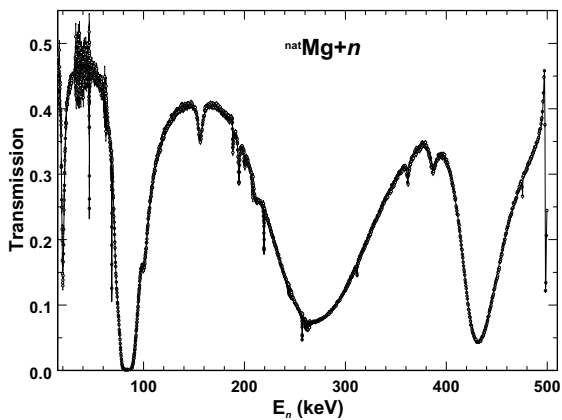}
\caption{\label{TransFig}$^{nat}$Mg+\textit{n} total
cross section data (points with error bars) from Ref. \protect\cite{We76}
(as transmissions) and SAMMY fit (solid curve).}
\end{figure}

\begin{figure}
\includegraphics*[height=150mm,width=150mm,keepaspectratio] {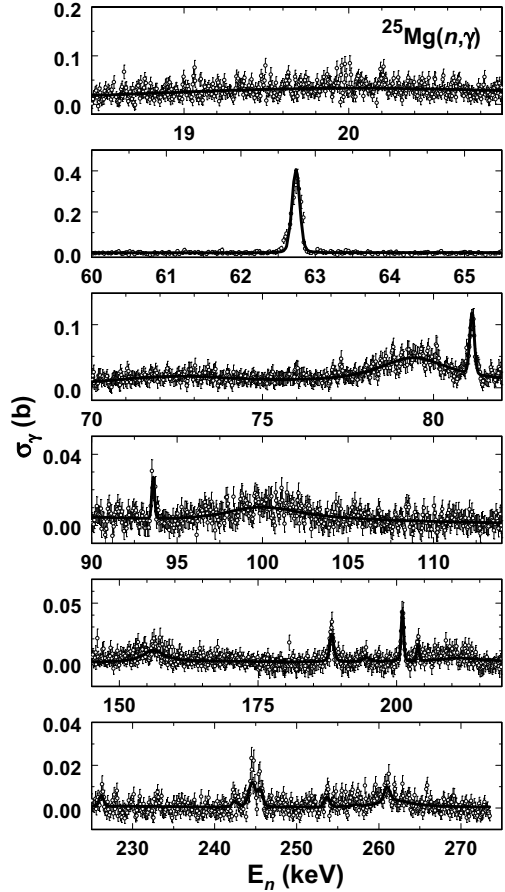}
\caption{\label{CapFig}$^{25}$Mg(\textit{n},$%
\protect\gamma $) cross section data \protect\cite{We76} (points with error
bars) and SAMMY fit (solid curves).}
\end{figure}

The starting parameters required considerable adjustments in some cases to
fit the data. Some of these differences probably can be ascribed to the
Breit-Wigner fitting approach of Ref. \cite{We76}. The resulting parameters
are given in Tables \ref{Mg24ResTable}, \ref{Mg26ResTable}, and \ref%
{Mg25ResTable}. All parameters for all observed resonances are included in
these tables, even those that are not well determined, so that the present
results could be duplicated if necessary. The $^{nat}$Mg+\textit{n} and $%
^{25}$Mg(\textit{n},$\gamma $)$^{26}$Mg data of Ref. \cite{We76} and the
SAMMY fits are shown in Figs. \ref{TransFig} and \ref{CapFig}, respectively.

\begin{table}[tbp] \centering%
\caption{$^{24}$Mg+\textit{n} resonance parameters. \label{Mg24ResTable}}%
\begin{tabular}{cccll}
\hline\hline
$E_{n}$ (keV) & \textit{l} & 2$J^{\pi }$ & $\Gamma _{\gamma }$ (eV) & g$%
\Gamma _{n}$ (eV) \\ \hline
\multicolumn{1}{l}{46.347$\pm $0.016} & (1) & (1$^{-}$) & 1.83\textbf{%
\footnotemark[1]%
} & 1.556$\pm $0.089 \\ 
\multicolumn{1}{l}{68.529$\pm $0.024} & (1) & (1$^{-}$) & 3\textbf{%
\footnotemark[2]%
} & 5.60$\pm $0.22 \\ 
\multicolumn{1}{l}{83.924$\pm $0.031} & 1 & 3$^{-}$ & 4.7\textbf{%
\footnotemark[1]%
} & 8007.0$\pm $5.0 \\ 
\multicolumn{1}{l}{176.700\textbf{%
\footnotemark[3]%
}} & (1) & (1$^{-}$) & 3\textbf{%
\footnotemark[2]%
} & 0.314\textbf{%
\footnotemark[4]%
} \\ 
\multicolumn{1}{l}{257.18$\pm $0.12} & (2) & (3$^{+}$) & 1.13\textbf{%
\footnotemark[1]%
} & 26.9$\pm $1.0 \\ 
\multicolumn{1}{l}{266.10$\pm $0.12} & 1 & 1$^{-}$ & 5.2\textbf{%
\footnotemark[1]%
} & 80216$\pm $41 \\ 
\multicolumn{1}{l}{431.07$\pm $0.23} & 1 & 3$^{-}$ & 7.0\textbf{%
\footnotemark[1]%
} & 30082$\pm $22 \\ 
\multicolumn{1}{l}{475.35$\pm $0.27} & 2 & 5$^{+}$ & 1.04\textbf{%
\footnotemark[1]%
} & 13.8$\pm $1.3 \\ 
\multicolumn{1}{l}{498.27$\pm $0.28} & 1 & 3$^{-}$ & 0.38\textbf{%
\footnotemark[1]%
} & 520.0$\pm $4.6 \\ \hline\hline
\end{tabular}%
\footnotetext[1]{Gamma width calculated to yield the corrected \cite{Ma81} capture kernel given in Ref. \cite{We76}.}
\footnotetext[2]{Assumed gamma width. See text for details.}
\footnotetext[3]{From Ref. \cite{We76}. Not observed in this work. See text for details.}
\footnotetext[4]{Neutron width calculated to yield the corrected \cite{Ma81} capture kernel given in Ref. \cite{We76}.}
\end{table}%

\begin{table}[tbp] \centering%
\caption{$^{26}$Mg+\textit{n} resonance parameters. \label{Mg26ResTable}}%
\begin{tabular}{cccll}
\hline\hline
$E_{n}$ (keV) & \textit{l} & 2$J^{\pi }$ & $\Gamma _{\gamma }$ (eV) & g$%
\Gamma _{n}$ (eV) \\ \hline
\multicolumn{1}{l}{68.7\textbf{%
\footnotemark[1]%
}} & (1) & (1$^{-}$) & 3\textbf{%
\footnotemark[2]%
} & 0.070\textbf{%
\footnotemark[3]%
} \\ 
\multicolumn{1}{l}{219.39$\pm $0.11} & 2 & 3$^{+}$ & 1.78\textbf{%
\footnotemark[4]%
} & 101.5$\pm $4.0 \\ 
\multicolumn{1}{l}{295.91$\pm $0.15} & 1 & 3$^{-}$ & 3\textbf{%
\footnotemark[2]%
} & 66920$\pm $170 \\ 
\multicolumn{1}{l}{427.38$\pm $0.25} & (0) & (1$^{+}$) & 4.3\textbf{%
\footnotemark[4]%
} & 3170$\pm $160 \\ 
\multicolumn{1}{l}{430.88$\pm $0.33} & \multicolumn{1}{l}{(1)} & 
\multicolumn{1}{l}{(1$^{-}$)} & 4.2\textbf{%
\footnotemark[4]%
} & 25990$\pm $290 \\ \hline\hline
\end{tabular}%
\footnotetext[1]{From Ref. \cite{We76}. Not observed in this work. See text for details.}
\footnotetext[2]{Assumed gamma width. See text for details.}
\footnotetext[3]{Neutron width calculated to yield the corrected \cite{Ma81} capture kernel given in Ref. \cite{We76}.}
\footnotetext[4]{Gamma width calculated to yield the corrected \cite{Ma81} capture kernel given in Ref. \cite{We76}.}
\end{table}%

\begingroup
\squeezetable
\begin{table*}[tbp] \centering%
\caption{$^{25}$Mg+\textit{n} resonance parameters. \label{Mg25ResTable}}%
\begin{tabular}{ccccccclllll}
\hline\hline
\multicolumn{5}{c}{$E_{n}$ (keV)} & \textit{l} & $J^{\pi }$ & $\Gamma
_{\gamma }$ (eV) & 2\textit{g}$\Gamma _{n}$ (eV) & \multicolumn{3}{c}{$%
\Gamma $ (eV)} \\ 
This Work & Ref. \cite{En98} & Ref. \cite{We76} & Ref. \cite{Ja2001} & Ref. %
\cite{Be69} &  &  &  &  & This Work & Ref. \cite{We76} & Ref. \cite{Ja2001}
\\ \hline
\multicolumn{1}{l}{19.880$\pm $0.014} & 19.7$\pm $0.2 & 19.90\textbf{%
\footnotemark[1]%
} &  &  & 0 & 2$^{+}$ & 1.732$\pm $0.031 & 2148$\pm $20 & 2580$\pm $24 &  & 
\\ 
& 51$\pm $6 &  &  &  &  &  &  &  &  &  &  \\ 
\multicolumn{1}{l}{62.738$\pm $0.023} & 62.5$\pm $0.2 & 62.88 & 60$\pm $10 & 
62.4 & 1\textbf{%
\footnotemark[2]%
} & 1$^{-}$\textbf{%
\footnotemark[2]%
} & 4.79$\pm $0.29 & 7.2$\pm $2.1 & 19.2$\pm $4.2 & 24.6$\pm $2.2 &  \\ 
\multicolumn{1}{l}{72.674$\pm $0.042} & 73.1$\pm $0.5 & 73.3 &  & 72.3 & 0 & 
2$^{+}$ & 4.56$\pm $0.29 & 3870$\pm $83 & 4650$\pm $100 & 7600$\pm $1100 & 
\\ 
\multicolumn{1}{l}{79.30$\pm $0.15} & 79.4$\pm $0.2 & 79.6 &  &  & 0 & 3$^{+}
$ & 6.17$\pm $0.24 & 2700$\pm $180 & 2320$\pm $150 & 1910$\pm $140 &  \\ 
\multicolumn{1}{l}{81.13$\pm $0.14} & 81.2$\pm $0.7 & 81.35 &  &  & (2) & (2$%
^{+}$) & 3\textbf{%
\footnotemark[3]%
} & 1.20$\pm $0.13 & $<$75 & 20.3$\pm $2.6 &  \\ 
\multicolumn{1}{l}{93.61$\pm $0.17} & 93.6$\pm $0.2 & 93.8 &  &  & (1) & (1$%
^{-}$) & 3\textbf{%
\footnotemark[3]%
} & 0.270$\pm $0.044 & $<$77 &  &  \\ 
\multicolumn{1}{l}{100.007$\pm $0.050} & 99.6$\pm $0.2 & 99.8 &  &  & 0 & 3$%
^{+}$ & 2.92$\pm $0.18 & 6074$\pm $85 & 5210$\pm $73 &  &  \\ 
& 102$\pm $2 &  &  &  &  &  &  &  &  &  &  \\ 
& 105.5$\pm $0.2 & 105.8 &  &  &  &  &  &  &  &  &  \\ 
156.169$\pm $0.076 & 156.3$\pm $0.2 & 156.5 &  &  & (1)\textbf{%
\footnotemark[6]%
} & 2$^{(-)}$ & 7.42$\pm $0.60 & 3759$\pm $89 & 4520$\pm $110 &  &  \\ 
188.334$\pm $0.081 & 188.6$\pm $0.2 & 188.9 &  &  & 0 & (2)$^{+}$ & 3.24$\pm 
$0.35 & 450$\pm $43 & 543$\pm $52 &  &  \\ 
194.502$\pm $0.085 & 194.0$\pm $0.2 & 194.2 &  &  & (1) & 4$^{(-)}$ & 0.59$%
\pm $0.24 & 2270$\pm $51 & 1514$\pm $34 &  &  \\ 
200.285$\pm $0.097 &  &  &  & 204 & 1 & 1$^{-}$\textbf{%
\footnotemark[4]%
} & 0.79$\pm $0.46 & 628$\pm $50 & 1257$\pm $100 &  &  \\ 
201.062$\pm $0.095 & 201.3$\pm $0.3 & 201.6 &  &  & (2) & (2$^{+}$)\textbf{%
\footnotemark[5]%
} & 4.26$\pm $0.60 & 10.7$\pm $5.0 & 17.1$\pm $6.0 &  &  \\ 
203.86$\pm $0.44 & 204.0$\pm $0.3 & 204.3 &  &  & (1) & (2$^{-}$) & 3\textbf{%
\footnotemark[3]%
} & 1.28$\pm $0.38 & $<$32 &  &  \\ 
211.20$\pm $0.11 & 209.8$\pm $0.5 & 210 &  &  & (1)\textbf{%
\footnotemark[6]%
} & (3$^{-}$) & 3.31$\pm $0.73 & 9400$\pm $140 & 8060$\pm $120 & 2630$\pm $%
230 &  \\ 
226.19$\pm $0.50 & 226.7$\pm $0.5 & 227 &  &  & (1) & (1$^{-}$) & 3\textbf{%
\footnotemark[3]%
} & 0.56$\pm $0.20 & $<$56 &  &  \\ 
242.45$\pm $0.55 &  &  &  &  & (1) & (1$^{-}$) & 3\textbf{%
\footnotemark[3]%
} & 0.30$\pm $0.16 & $<$43 &  &  \\ 
244.58$\pm $0.12 & 244.7$\pm $0.5 & 245 & 235$\pm $2 & 250 & 1\textbf{%
\footnotemark[7]%
} & 1$^{-}$ & 3.63$\pm $0.47 & 212$\pm $43 & 428$\pm $86 &  & 250$\pm $170
\\ 
245.57$\pm $0.56 & 245$\pm $2 &  &  &  & (1) & (1$^{-}$) & 3\textbf{%
\footnotemark[3]%
} & 1.40$\pm $0.50 & $<$34 &  &  \\ 
253.67$\pm $0.58 &  &  &  &  & (1) & (1$^{-}$) & 3\textbf{%
\footnotemark[3]%
} & 0.71$\pm $0.28 & $<$48 &  &  \\ 
261.00$\pm $0.14 & 260.7$\pm $0.5 & 261 &  &  & 0 & (2$^{+}$) & 1.18$\pm $%
0.27 & 128$\pm $35 & 155$\pm $42 &  &  \\ 
261.9$\pm $0.14 &  &  &  &  & (1) & 4$^{(-)}$ & 1.82$\pm $0.38 & 6200$\pm $%
280 & 4140$\pm $190 &  &  \\ 
& 279.8$\pm $0.6 &  &  &  &  &  &  &  &  &  &  \\ 
& 282.8$\pm $0.6 &  &  &  &  &  &  &  &  &  &  \\ 
& 290.7$\pm $0.6 &  &  &  &  &  &  &  &  &  &  \\ 
311.56$\pm $0.15 & 311.7$\pm $0.6 &  &  &  & (2) & 5$^{(+)}$ & 3\textbf{%
\footnotemark[3]%
} & 532$\pm $35 & 293$\pm $19 &  &  \\ 
& 345.7$\pm $0.7 &  &  &  &  &  &  &  &  &  &  \\ 
361.88$\pm $0.19 & 360.7$\pm $0.7 &  & 362$\pm $2 &  & 2\textbf{%
\footnotemark[7]%
} & 4$^{+}$ & 3\textbf{%
\footnotemark[3]%
} & 2200$\pm $120 & 1470$\pm $80 &  & 2100$\pm $900 \\ 
& 379$\pm $2 &  &  &  &  &  &  &  &  &  &  \\ 
387.35$\pm $0.21 & 385.8$\pm $0.8 &  & 383$\pm $2 &  & 3\textbf{%
\footnotemark[7]%
} & 5$^{-}$ & 3\textbf{%
\footnotemark[3]%
} & 12000$\pm $160 & 6550$\pm $87 &  & 9300$\pm $2500 \\ \hline\hline
\end{tabular}%
\footnotetext[1]{Attributed to Ref. \cite{Si74} in Ref. \cite{We76}.}
\footnotetext[2]{Spin and parity assignment based on Ref. \cite{Be69}.}
\footnotetext[3]{Assumed gamma width. See text for details.}
\footnotetext[4]{Parity assignment based on Ref. \cite{Be69}.}
\footnotetext[5]{\textit{J}>1.}
\footnotetext[6]{\textit{l}>0.}
\footnotetext[7]{Assigned natural parity because observed in $^{22}Ne(\alpha,n)^{25}$Mg.}
\end{table*}
\endgroup%

\bigskip

Although I did not have $^{24,26}$Mg(\textit{n},$\gamma $)$^{25,27}$Mg data
to fit, the gamma widths for the strong resonances and the neutron widths
for the weak resonances presented herein were calculated to be consistent
with the corrected \cite{Ma81} neutron capture data of Ref. \cite{We76}. For
strong resonances that were clearly visible in the total cross section data,
the $\Gamma _{\gamma }$ values in Tables \ref{Mg24ResTable} and \ref%
{Mg26ResTable} were calculated to yield the corrected \cite{Ma81} capture
kernels ($g\Gamma _{n}\Gamma _{\gamma }/\Gamma $) given in Ref. \cite{We76}.
Because $\Gamma _{n}>>\Gamma _{\gamma }$ for these resonances, the choice of 
$\Gamma _{\gamma }$ has negligible effect on the fit to the total cross
section data. For weak resonances not visible in the total cross section, I
used $\Gamma _{\gamma }$ = 3.0 eV and the corrected \cite{Ma81} capture
kernels of Ref. \cite{We76} to calculate the neutron widths. I used $\Gamma
_{\gamma }$ = 3.0 eV because it appears to be close to the average gamma
width for these nuclides. In these cases, it is clear that the neutron
widths are fairly small. I also used $\Gamma _{\gamma }$ = 3.0 eV for those
cases where neither capture kernels nor $\Gamma _{\gamma }$ values were
given in Refs. \cite{We76,Mu81}. Because the neutron widths\ for these
resonances are large, any physically reasonable choice of $\Gamma _{\gamma }$
could be used to fit the total cross section data. I also used $\Gamma
_{\gamma }$ = 3.0 eV in those cases where resonances were visible only in
the $^{25}$Mg(\textit{n},$\gamma $)$^{26}$Mg data and fitted the data to
obtain the neutron widths. In these cases, only the capture kernels are well
determined, so the individual $\Gamma _{\gamma }$ and 2\textit{g}$\Gamma
_{n} $ values given in Table \ref{Mg25ResTable} are rather arbitrary.
However, the widths of peaks in the neutron capture data and/or the total
cross section data can be used to set limits on neutron, and hence the
total, widths in these cases. For these cases, the tenth column in Table \ref%
{Mg25ResTable} lists the limits on the total widths rather than the actual
total widths ($\Gamma _{\gamma }+\Gamma _{n}$) used to fit the data.

One-standard-deviation uncertainties in the partial widths determined in
fitting the data are also given in Tables \ref{Mg24ResTable}, \ref%
{Mg26ResTable}, and \ref{Mg25ResTable}. Uncertainties in the partial widths
were added in quadrature to obtain uncertainties in the total widths.
Uncertainties in the resonance energies are dominated by the flight path
length uncertainty, $\Delta $d$\approx $3 cm. Because the flight path length
for neutron capture was shorter than that for the total cross section
measurements, energy uncertainties for resonances observed only in the
neutron capture data are correspondingly larger. Weak and very broad
resonances can also have additional nonnegligible uncertainties in their
energy associated with the fitting process. The two uncertainties were added
in quadrature to obtain the values given in Tables \ref{Mg24ResTable}, \ref%
{Mg26ResTable}, and \ref{Mg25ResTable}.

If the neutron width of a resonance is large enough, then it is possible to
discern its spin and/or parity. For example, \textit{s}-wave resonances have
a characteristic asymmetric shape in the total cross section due to
interference with potential scattering. On this basis, the $^{25}$Mg+\textit{%
n} resonances at $E_{n}$=19.880, 72.674, 79.30, 100.007, 188.334, and 261.00
keV definitely can be assigned as being \textit{s} wave. Similarly, although
it is not possible to discern the parity of the $^{25}$Mg+\textit{n}
resonances at $E_{n}$=156.169 and 211.20 keV, they are definitely not 
\textit{s} wave. In addition, the spins of the $^{25}$Mg+\textit{n}
resonances at $E_{n}$=19.880, 72.674, 79.30, 100.007, 156.169, 194.502,
200.285, 244.58, 261.90, 311.56, 361.88, and 387.35 keV definitely can be
assigned by virtue of the heights of the peaks in the total cross section
(depths of the dips in the transmission spectrum).

For the present application, it is important to identify natural parity
resonances in $^{25}$Mg+\textit{n} because only they can participate in the $%
^{22}$Ne($\alpha $,\textit{n})$^{25}$Mg reaction. On the basis of the
neutron data alone, there are at least 16 states in $^{26}$Mg between
threshold and the lowest observed $^{22}$Ne($\alpha $,\textit{n})$^{25}$Mg
resonance, two of which ($E_{n}$=19.880 and 72.674 keV) definitely have
natural parity and two others ($E_{n}$=79.30 and 100.007 keV) definitely
have non-natural parity. From the present analysis, together with
information from $^{26}$Mg($\gamma $,\textit{n})$^{25}$Mg measurements \cite%
{Be69}, two more ($E_{n}$=62.738 and 200.285 keV) of the states in this
region can be assigned as natural parity. In the next section, this and
other issues arising from comparisons to previous work are discussed.

\section{Comparison to Previous Work\label{CompSec}}

The $^{24,26}$Mg+\textit{n} resonance parameters of the present work are,
with a few notable exceptions, in agreement with those of Refs. \cite%
{We76,Ma81,Mu81} to within the experimental uncertainties. Exceptions
include the $^{24}$Mg+\textit{n} resonance at 46.347 keV which was
previously assigned as a definite $\frac{3}{2}^{+}$ \cite{Mu81}. This
resonance is so weak in the total cross section that it was not possible to
make a firm $J^{\pi }$ assignment in the present work. In addition, I find
that the width of the $^{24}$Mg+\textit{n} resonance at 498.27 keV is almost
10 standard deviations larger than given in Refs. \cite{We76,Mu81}. Also,
the neutron width (and hence the total width) of the $^{26}$Mg+\textit{n}
resonance at 219.39 keV was found to be four times smaller in the present
work than that given in Refs. \cite{We76,Mu81}. Also, a broad $\frac{1}{2}%
^{(-)}$ $^{26}$Mg+\textit{n} resonance at $E_{n}$=300$\pm $4 keV is listed
in Ref. \cite{Mu81} but not in Ref. \cite{We76}. I find that the fit to the
total cross section data is much improved if a $^{26}$Mg+\textit{n}
resonance is included with an energy and width in agreement with those give
in Ref. \cite{Mu81}, but only if $J^{\pi }$=$\frac{3}{2}^{-}$. In addition,
the 68.529-keV resonance listed in Table \ref{Mg24ResTable} has not been
noted in any previous study but is clearly visible in the total cross
section data. This resonance should be visible in the isotopic (\textit{n},$%
\gamma $) data, but the data are not available in this energy range for $%
^{25}$Mg, or at all for $^{24,26}$Mg, so I tentatively assign this resonance
to $^{24}$Mg+\textit{n}. The neutron width is clearly too large to
correspond to the 68.7-keV resonance attributed to $^{26}$Mg+\textit{n} in
Ref. \cite{We76}. Finally, this latter resonance as well as the 176.7-keV
resonance in $^{24}$Mg+\textit{n} were not visible in the total cross
section, but they are included in Tables \ref{Mg24ResTable} and \ref%
{Mg26ResTable} in the interest of completeness.

The $^{25}$Mg+\textit{n} parameters extracted in the present work are
compared to parameters resulting from a previous analysis of these same data %
\cite{We76} as well as to information from $^{26}$Mg($\gamma $,\textit{n})$%
^{25}$Mg \cite{Be69} and $^{22}$Ne($\alpha $,\textit{n})$^{25}$Mg \cite%
{Ja2001} measurements and to a recent compilation \cite{En98} in Table \ref%
{Mg25ResTable}. Unless otherwise noted, the parameters in Table \ref%
{Mg25ResTable} are from the present work. To aid the comparison between the
various experiments, the energies of Refs. \cite{Be69,En98,Ja2001} have been
converted to laboratory neutron energies using the Q-values given in Ref. %
\cite{En98}.

Overall, there is fairly good agreement between the results of the present
work and previous studies although, except for excitation energies, there is
sparse information about states in $^{26}$Mg in this energy range from
previous work. In order of increasing energy, important correspondences with
and differences between the present and previous work are outlined in the
next several paragraphs.

\subsection{Neutron resonances below the lowest energy $^{22}$Ne($\protect%
\alpha $,\textit{n})$^{25}$Mg resonance}

One of only two definite J$^{\pi }$ assignments in this energy range was
made in Ref. \cite{Si74} at $E_{n}$=19.9 keV. This assignment is confirmed
in the present work although the width needed to fit the data is
considerably larger than given in Ref. \cite{Si74}.

A natural-parity state at $E_{x}$=11142 keV ($E_{n}$=51 keV) tentatively
assigned in Ref. \cite{En98} was not observed in this work. This state has
been shown \cite{Dr93} to be an erroneous assignment \cite{Dr91, Ha91} due
to background from the $^{11}$B($\alpha $,\textit{n})$^{14}$N reaction.

Resonances at $E_{n}$=62.738 and 72.674 keV correspond well with the $E_{L}$%
=54.3- and 63.2-keV ($E_{n}$=62.4 and 72.3 keV according to equations 7 and
8 in Ref. \cite{Bo67}) resonances observed in the $^{26}$Mg($\gamma $,%
\textit{n})$^{25}$Mg \cite{Be69} reaction. Note that the neutron energies $%
E_{n}$, corresponding to the $E_{L}$ values of Ref. \cite{Be69}, given in
the footnote of Table 3 of Ref. \cite{Wo89} are incorrect. The former
resonance was assigned as $J^{\pi }$=1$^{-}$ in Ref. \cite{Be69} and is a
strong resonance in $^{25}$Mg(\textit{n},$\gamma $)$^{26}$Mg but because it
is barely visible in the $^{nat}$Mg+\textit{n} total cross section data, it
is not possible to make a firm $J^{\pi }$ assignment based on the data of
Ref. \cite{We76}. The total width fitted in the present work is in agreement
with Ref. \cite{We76} but the capture kernel I obtained is 30\% smaller. The 
$J^{\pi }$ value of the $E_{L}$=63.2-keV resonance is not discussed in Ref. %
\cite{Be69}, but the firm 2$^{+}$ assignment from the present work is
consistent with the small size of the peak in the $^{26}$Mg($\gamma $,%
\textit{n})$^{25}$Mg data.

The precision of the width of the $E_{n}$=81.13-keV resonance given in Ref. %
\cite{We76} seems insupportable. This resonance could not be observed in the
total cross section due to a nearby broad $^{24}$Mg+\textit{n} resonance;
hence, only the area and width of the peak in the $^{25}$Mg(\textit{n},$%
\gamma $)$^{26}$Mg data can be used to determine the resonance parameters.
The resolution of the experiment at this energy was 120 eV, or five times
the width of the resonance assigned in Ref. \cite{We76}. I found that the
data could be well fitted with widths as large as 75 eV. Because the data
could be fitted by such a wide range of partial widths, I decided to hold $%
\Gamma _{\gamma }$ fixed at 3 eV and vary only $\Gamma _{n}$ while fitting
the data.

The state in Ref. \cite{En98} at $E_{x}$=11191 keV ($E_{n}$=102 keV) was not
observed in this work, but its existence can not be ruled out due to the
presence of the broad resonance at $E_{n}$=100.007 keV.

A state at $E_{x}$=11194.5 keV ($E_{n}$=105.5 keV) is listed as a firm $%
J^{\pi }$=2$^{+}$ assignment having a fairly large width ($\Gamma $=10$\pm $%
2 keV) in Refs. \cite{Mu81,En98} and apparently is based on the work of Ref. %
\cite{We76} in which a weak resonance was assigned at $E_{n}$=105.8 keV,
although no width is given in this latter reference. Such a broad resonance
easily would be visible in the total cross section data analyzed in this
work, but there is no sign of it. Perhaps the compilers have confused it
with the broad resonance at $E_{n}$=100.007 keV. Although it is possible to
add a small resonance at $E_{n}$=105.8 keV, the fit to the data is not
improved by its inclusion.

A doublet is required to fit the data near $E_{n}$=201 keV rather than the
single resonance listed in previous work. The lower resonance in this pair
has a much larger neutron width than the upper one and definitely can be
assigned as \textit{J}=1. It appears to correspond to the $E_{L}$=182 keV ($%
E_{n}$=204 keV) resonance in $^{26}$Mg($\gamma $,\textit{n})$^{25}$Mg \cite%
{Be69} and so is assigned natural parity in Table \ref{Mg25ResTable}.

The width of the resonance at $E_{n}$=211.20 keV fitted in this work is
three times larger than determined in Ref. \cite{We76} and reported in Refs. %
\cite{Mu81,En98}. Although it is clear that there is a broad $^{25}$Mg+%
\textit{n} resonance at this energy, the data could not be fitted as well as
one would like with any single resonance, although it is clear a broad 
\textit{s}-wave resonance is ruled out. There is also a fairly weak
resonance just below the 211.20-keV resonance visible in the total cross
section data. Because there is no sign of this resonance in the $^{25}$Mg(%
\textit{n},$\gamma $)$^{26}$Mg data, it is likely to be due to one of the
other two Mg isotopes.

\subsection{Which neutron resonance corresponds to the lowest energy $^{22}$%
Ne($\protect\alpha $,\textit{n})$^{25}$Mg resonance?}

There are four neutron resonances ($E_{n}$=226.19, 242.45, 244.58, and
245.57 keV) near the energy ($E_{n}$=235$\pm $2 keV) corresponding to the
lowest observed $^{22}$Ne($\alpha $,\textit{n})$^{25}$Mg resonance. None of
these four resonances has an energy in agreement with the reported $^{22}$Ne(%
$\alpha $,\textit{n})$^{25}$Mg resonance to within the experimental
uncertainties, but only one ($E_{n}$=244.58 keV) is broad enough to
corresponed to the width reported in the latest $^{22}$Ne($\alpha $,\textit{n%
})$^{25}$Mg measurements \cite{Ja2001}. It appears that either the width or
the energy reported in Ref. \cite{Ja2001} is in error, and if the reported
width is correct, then the partial widths determined in this work indicate
that different resonances near this energy have been observed in the $^{22}$%
Ne($\alpha $,\textit{n})$^{25}$Mg and $^{22}$Ne($\alpha $,$\gamma $)$^{26}$%
Mg \ reactions.

The first two neutron resonances in this region ($E_{n}$=226.19 and 242.45
keV) are visible only as small peaks in the $^{25}$Mg(\textit{n},$\gamma $)$%
^{26}$Mg data; hence, they have relatively small neutron widths. The
higher-energy one has not been reported in any previous work. The upper two
resonances in this region ($E_{n}$=244.58, and 245.57 keV) appear as a
partially resolved doublet in the $^{25}$Mg(\textit{n},$\gamma $)$^{26}$Mg
data. Only the lower-energy one of this pair is visible in the $^{nat}$Mg+%
\textit{n} total cross section data, via which it definitely can be assigned 
\textit{J}=1.

In addition to the previous analysis of these data \cite{We76} and the $%
^{22} $Ne($\alpha $,\textit{n})$^{25}$Mg \cite{Dr91, Ha91,Dr93,Ja2001} work,
resonances near this energy have been identified in $^{26}$Mg($\gamma $,%
\textit{n})$^{25}$Mg \cite{Be69}, $^{22}$Ne($\alpha $,$\gamma $)$^{26}$Mg %
\cite{Wo89}, and $^{22}$Ne($^{6}$Li,\textit{d})$^{26}$Mg \cite{Gi93}
measurements. Data from the latter two and $^{22}$Ne($\alpha $,\textit{n})$%
^{25}$Mg reactions have been interpreted as having observed the same state
in $^{26}$Mg corresponding to an $E_{\alpha }\approx $830-keV resonance. If
the fairly large width reported in Ref. \cite{Ja2001} is correct, then the
bulk of the width must be due to the neutron channel and a corresponding
resonance easily would be visible in the data of Ref. \cite{We76}. In this
case, the only possible corresponding resonance is at $E_{n}$=244.58 keV.
The energy of this resonance is almost 5 standard deviations higher than the
energy determined in Ref. \cite{Ja2001}, but its width is in good agreement
with Ref. \cite{Ja2001} whereas all the other neutron resonances near this
energy are too narrow. Given the much superior energy resolution of the data
of Ref. \cite{We76}, the excellent correspondence of the energies determined
from the $^{nat}$Mg+\textit{n} total cross section and $^{25}$Mg(\textit{n},$%
\gamma $)$^{26}$Mg data, and the vast quantity of data on other nuclides
taken with this apparatus, it is extremely unlikely that the energy of the $%
E_{n}$=244.58-keV resonance could be in error by such a large amount.
Therefore, if the width in Ref. \cite{Ja2001} is correct, then the reported
energy must be almost 10 keV too low. Furthermore, if the width for the $%
E_{\alpha }$=832-keV resonance reported in Ref. \cite{Ja2001} is correct,
then the parameters reported in that work and Ref. \cite{Wo89} indicate that
different resonances were observed in the $^{22}$Ne($\alpha $,$\gamma $)$%
^{26}$Mg and $^{22}$Ne($\alpha $,\textit{n})$^{25}$Mg measurements, and the
partial widths determined in this work indicate that the resonance observed
in the $^{22}$Ne($\alpha $,$\gamma $)$^{26}$Mg measurements would not be
seen in the $^{22}$Ne($\alpha $,\textit{n})$^{25}$Mg measurements and vice
versa.

The width ($\Gamma $=250$\pm $170 eV) and strength ($\omega \gamma _{(\alpha
,n)}=(2J+1)\Gamma _{\alpha }\Gamma _{n}/\Gamma =118\pm 11$ $\mu $eV) for the 
$E_{\alpha }$=832$\pm $2-keV resonance from the $^{22}$Ne($\alpha $,\textit{n%
})$^{25}$Mg measurements \cite{Ja2001}, together with the strength ($\omega
\gamma _{(\alpha ,\gamma )}=(2J+1)\Gamma _{\alpha }\Gamma _{\gamma }/\Gamma
=36\pm 4$ $\mu $eV) of the $E_{\alpha }$=828$\pm $5-keV resonance from the $%
^{22}$Ne($\alpha $,$\gamma $)$^{26}$Mg measurements \cite{Wo89} imply $%
\Gamma _{\gamma }$=58 eV if the same resonance is being observed in both
reactions. This is almost 20 times larger than the average radiation width
for $^{26}$Mg and, although radiation widths vary more widely in this mass
range than for heavier nuclides, it is considerably larger than any reported
radiation width for nuclides in this mass range. Furthermore, the partial
widths resulting from assuming the same resonance has been observed in both
the ($\alpha $,\textit{n}) and ($\alpha $,$\gamma $) channels imply a
capture kernel ($g\Gamma _{n}\Gamma _{\gamma }/\Gamma $) roughly ten times
larger than observed for any of the four $^{25}$Mg(\textit{n},$\gamma $)$%
^{26}$Mg resonances near this energy.

If instead different resonances were observed in the ($\alpha $,\textit{n})
and ($\alpha $,$\gamma $) reactions, then the strength of the resonance
observed in $^{22}$Ne($\alpha $,\textit{n})$^{25}$Mg together with the
partial widths for the $E_{n}$=244.58-keV resonance from the present work ($%
\Gamma _{n}/\Gamma _{\gamma }=117$) imply a resonance strength in the $^{22}$%
Ne($\alpha $,$\gamma $)$^{26}$Mg reaction of $\omega \gamma _{(\alpha
,\gamma )}$=1.0 $\mu $eV, well below the sensitivity of the measurements of
Ref. \cite{Wo89}. Similarly, if the $E_{\alpha }$=828-keV resonance from $%
^{22}$Ne($\alpha $,$\gamma $)$^{26}$Mg is identified with either the $E_{n}$%
=226.19- or 242.45-keV resonance from the present work, then the strength
from the $^{22}$Ne($\alpha $,$\gamma $)$^{26}$Mg measurements together with
the partial widths from the present work imply a strength in the $^{22}$Ne($%
\alpha $,\textit{n})$^{25}$Mg reaction of about $\omega \gamma _{(\alpha ,n)}
$=10 $\mu $eV. This is smaller than any resonance reported in Ref. \cite%
{Ja2001}, and in any case would have been obscured by the much stronger $%
E_{\alpha }$=832-keV resonance if indeed different resonances near this
energy were being observed in the $^{22}$Ne($\alpha $,\textit{n})$^{25}$Mg
and $^{22}$Ne($\alpha $,$\gamma $)$^{26}$Mg reactions.

Alternatively, if the width reported in Ref. \cite{Ja2001} for the $%
E_{\alpha }$=832$\pm $2-keV resonance is too large, then it is possible that
the same state in $^{26}$Mg has been observed in the various reactions and
the corresponding neutron resonance is at $E_{n}$=226.19 or 242.45 keV.
However, the energies still do not agree to within the reported experimental
uncertainties.

Interestingly, a resonance at $E_{L}$=224 keV, corresponding to $E_{n}$=250
keV was observed \cite{Be69} in the $^{26}$Mg($\gamma $,\textit{n})$^{25}$Mg
reaction and assigned $J$=1 in agreement with the $E_{n}$=244.58-keV
resonance of the present work, but it was tentatively assigned as being
non-natural parity. Although no energy uncertainties were stated in Ref. %
\cite{Be69}, information given in Ref. \cite{Bo67} implies that the energy
of the $E_{L}$=224-keV resonance corresponds to the $E_{n}$=244.58-keV
resonance observed in this work to well within the experimental
uncertainties. Also, a state in $^{26}$Mg at $E_{x}$=11311$\pm $20 keV
(corresponding to $E_{\alpha }$=828 or $E_{n}$=231 keV), having a
spectroscopic factor $S_{\alpha }$=0.04, was observed in $^{22}$Ne($^{6}$Li,%
\textit{d})$^{26}$Mg measurements \cite{Gi93} and was assigned $J^{\pi }$=2$%
^{+}$ although 1$^{-}$ could not be ruled out. Because the energy resolution
of the experiment was fairly broad ($\Delta E$=120 keV), these measurements
are not helpful in ascertaining whether there is more than one $^{22}$Ne+$%
\alpha $ resonance near this energy.

\subsection{Higher energy resonances}

The possible multiplet identified in Ref. \cite{We76} near $E_{n}$=261 keV
clearly shows up as a doublet in both the $^{25}$Mg(\textit{n},$\gamma $)$%
^{26}$Mg and $^{nat}$Mg+\textit{n} total cross section data. The
lower-energy resonance of the pair is clearly \textit{s}-wave whereas the
upper-energy one can be assigned as a definite $J$=4 resonance.

The lack of $^{25}$Mg(\textit{n},$\gamma $)$^{26}$Mg data was a severe
handicap to the analysis above this energy. However, the analysis was
continued to $E_{n}$=500 keV in an attempt to overlap with the $^{22}$Ne($%
\alpha $,\textit{n})$^{25}$Mg data as much as possible. Of the eight
resonances listed in Ref. \cite{En98} in this region, three were observed,
two of which correspond to $^{22}$Ne($\alpha $,\textit{n})$^{25}$Mg
resonances.

The second and third lowest energy $^{22}$Ne($\alpha $,\textit{n})$^{25}$Mg
resonances \cite{Ja2001} at $E_{\alpha }$= 976, and 1000 keV appear to
correspond to the $E_{n}$= 361.88-, and 387.35-keV resonances, respectively,
observed in this work. There is fairly good agreement in the widths and
energies between the neutron and $^{22}$Ne($\alpha $,\textit{n})$^{25}$Mg
data. Curiously, the $J$-values required to fit the neutron data imply
rather high \textit{l}$_{\alpha }$ values for these two resonances. Because
the $E_{n}$=387.35-keV resonance is also observed as a resonance at $%
E_{\alpha }$=1000 keV in the $^{22}$Ne($\alpha $,\textit{n})$^{25}$Mg
reaction, it is assigned as natural parity (\textit{l}$_{n}$=3) in Table \ref%
{Mg25ResTable} even though \textit{f}-waves were not included in the $%
\mathcal{R}$-matrix analysis. It should make little difference to the
quality of the fit to the data that the fitted cross section was calculated
with a \textit{d}-wave rather than an \textit{f}-wave resonance at this
energy.

\section{Impact on the $^{22}$N\lowercase{e} ($\protect\alpha $,\textit{
\lowercase{n}})$^{25}$M\lowercase{g} astrophysical reaction rate\label%
{ReacRate}}

At \textit{s}-process temperatures, the uncertainty in the $^{22}$Ne($\alpha 
$,\textit{n})$^{25}$Mg reaction rate is dominated by possible contributions
from undetected resonances below the lowest observed resonance at $E_{\alpha
}$=832 keV. Most previous evaluations of this rate have assumed either that
all the known states, or at most only a single state, in $^{26}$Mg in this
energy range can contribute to the reaction rate. To contribute to this
reaction rate, the state must have natural parity. Two \cite{Mu81} or three %
\cite{En98} states in this region have been assigned natural parity in the
compilations. As discussed above, two of these states do not even exist, let
alone have natural parity. The third ($E_{x}$=11112.18, $E_{n}$=19.880 keV)
was identified via the neutron total cross section measurements of Ref. \cite%
{Si74} and this assignment was verified in the present work. In addition,
another state ($E_{x}$=11162.95, $E_{n}$=72.674 keV) has been assigned
definite natural parity ($J^{\pi }$=2$^{+}$) in the present work, two others
($E_{x}$=11169.32, 11189.23, $E_{n}$=79.30, 100.007 keV) have been assigned
definite non-natural parity ($J^{\pi }$=3$^{+}$), and together with
information from $^{26}$Mg($\gamma $,\textit{n})$^{25}$Mg measurements \cite%
{Be69}, two more ($E_{x}$=11153.40, 11285.65, $E_{n}$=62.738, 200.285 keV)
can be assigned as very likely natural parity ($J^{\pi }$=1$^{-}$).
Therefore, at the very least four states of the 16 (17 if the $E_{x}$=11191, 
$E_{n}$=102 keV state of Ref. \cite{En98} is included) in this energy range
should be included when estimating the contributions of possible low-energy
resonances to the $^{22}$Ne($\alpha $,\textit{n})$^{25}$Mg reaction rate,
and two states definitely can be eliminated from consideration.

The yields below the lowest observed $^{22}$Ne($\alpha $,\textit{n})$^{25}$%
Mg resonance together with the quoted upper limit on the strength of the
possible $E_{\alpha }$=635-keV resonance from Ref. \cite{Ja2001} can be used
to estimate the contributions of this and other possible unobserved
resonances to the $^{22}$Ne($\alpha $,\textit{n})$^{25}$Mg reaction rate.
Assuming that the thick target approximation applies \cite{Ro88}, the upper
limit on the resonance strength ($\omega \gamma _{2}$) of a possible
resonance at energy $E_{2}$ having yield $Y_{2}$ scales from the measured
limits on the strength ($\omega \gamma _{1}$) and yield ($Y_{1}$) of the $%
E_{\alpha }$=$E_{1}$=635-keV resonance as:%
\begin{equation}
\omega \gamma _{_{2}}=\omega \gamma _{1}\frac{Y_{2}E_{2}}{Y_{1}E_{1}}.
\label{StrengthEstimae}
\end{equation}

The yield limits of Ref. \cite{Ja2001} in this energy are approximately
constant, so I simply scaled the strengths of possible low-energy resonances
from the measured limit for the $E_{\alpha }$=635-keV resonance according to
their energies. The contribution of these resonances to the $^{22}$Ne($%
\alpha $,\textit{n})$^{25}$Mg reaction rate then can be estimated using the
simple $\delta $-resonance formula \cite{Ba69}:%
\begin{equation}
N_{A}<\sigma v>_{r}=1.54\times 10^{5}A^{-3/2}T_{9}^{-3/2}(\omega \gamma
)e^{-11.605E_{r}/T_{9}}.  \label{ResReacRate}
\end{equation}

Here $A$ is the reduced mass, $T_{9}$ is the temperature in GK, $(\omega
\gamma )_{r}$ is the resonance strength in eV, $E_{r}$ is the c.m. resonance
energy in MeV, and $N_{A}<\sigma v>$ is the reaction rate in cm$^{3}$%
/s/mole. However, in some cases the widths of the resonances may be
important, so I also calculated the reaction rate by numerically integrating
the cross section calculated from the resonance parameters. To do this, I
used the definitions of the resonance strength ($\omega \gamma _{(\alpha
,n)}=(2J+1)\Gamma _{\alpha }\Gamma _{n}/\Gamma $) and total width ($\Gamma
=\Gamma _{n}+\Gamma _{\gamma }+\Gamma _{\alpha }$) to calculate the alpha
widths ($\Gamma _{\alpha }=\frac{\omega \gamma (\Gamma _{n}+\Gamma _{\gamma
})}{(2J+1)\Gamma _{n}-\omega \gamma }$) from the scaled resonance strengths
as well as the partial widths determined in the present work. Then, I used
these partial widths in SAMMY to calculate the $^{25}$Mg(\textit{n},$\alpha $%
)$^{22}$Ne \ cross section, from which the $^{22}$Ne($\alpha $,\textit{n})$%
^{25}$Mg cross section was calculated using detailed balance. The two
approaches were in satisfactory agreement for the purposes of the present
work. For example, using the measured \cite{Ja2001} upper limit for the
strength of the $E_{\alpha }$=635-keV resonance of 60 neV and the $J^{\pi }$%
, $\Gamma _{n}$, and $\Gamma _{\gamma }$ values for the $E_{n}$=62.738-keV
resonance in Table \ref{Mg25ResTable} leads to $\Gamma _{\alpha }$$<$27 
neV, and a reaction rate at $T_{9}$=0.1 of 3.3$\times $10$%
^{-29}$ cm$^{3}$/s/mole from numerical integration and 3.4$\times $10$^{-29}$
cm$^{3}$/s/mole from the $\delta $-resonance formula.

\begin{figure}
\includegraphics* [width=85mm,height=85mm,keepaspectratio]{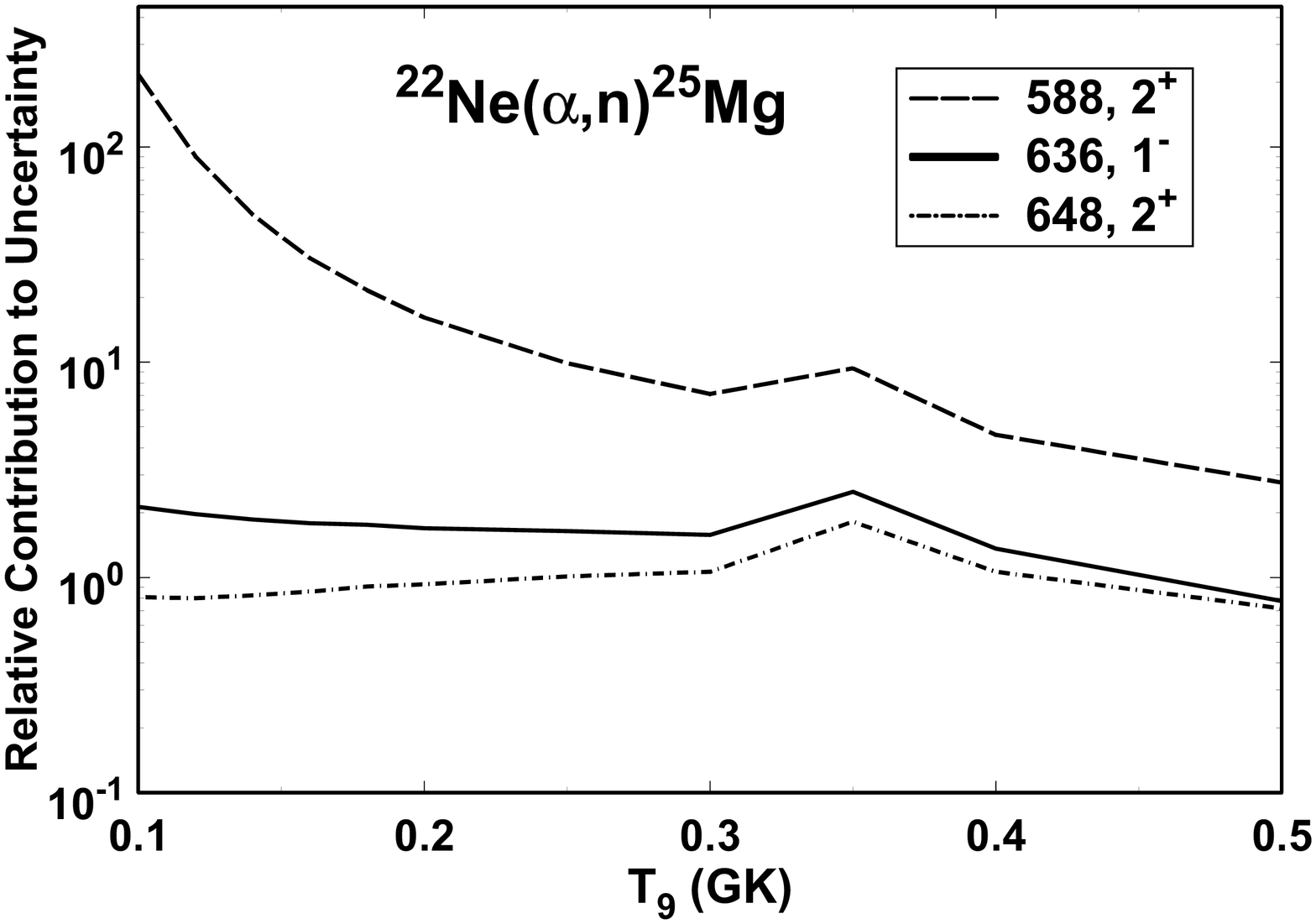}
\caption{\label{RatioFig}Ratios of the individual
contributions of three possible resonances (labeled by their laboratory
alpha-particle energies and J$^{\protect\pi }$ values) to the $^{22}$Ne($%
\protect\alpha $,\textit{n})$^{25}$Mg reaction rate to the uncertainty
(``High''-``Recomm.'') of Ref. \protect\cite{Ja2001} versus temperature.}
\end{figure}

The contributions of the two definite natural parity ($J^{\pi }$=2$^{+}$, $%
E_{n}$=19.880 and 72.674 keV) states and one very likely natural parity
state ($J^{\pi }$=1$^{-}$, $E_{n}$=62.738 keV) to the uncertainty in the
reaction rate are shown in Fig. \ref{RatioFig}. Shown in this figure are the
reaction rates due to each of the resonances divided by the difference
between the ``High'' and ``Recomm.'' rates of Ref. \cite{Ja2001}. The other
very likely natural parity state ($J^{\pi }$=1$^{-}$, $E_{n}$=200.285 keV)
contributes much less to the uncertainty so it is not shown. As can be seen
in Fig. \ref{RatioFig}, the present results indicate that the uncertainty in
the reaction rate calculated in Ref. \cite{Ja2001} is approximately a factor
of 10 too small at \textit{s}-process temperatures. Most of the increase in
the uncertainty indicated by the present work results from inclusion of the $%
E_{n}$=19.880-keV resonance. The natural-parity nature of this state has
been known for many years \cite{Si74}, but it often has been overlooked when
estimating the uncertainty in the $^{22}$Ne($\alpha $,\textit{n})$^{25}$Mg
reaction rate. Instead, most of the attention has been focussed on the $%
E_{n} $=62.738-keV resonance since attention was first called to it in Ref. %
\cite{Be69}. The contribution of the $E_{n}$=62.738-keV resonance to the
reaction rate uncertainty appears to have been underestimated by a factor of
two in Ref. \cite{Ja2001}. Both the $\delta $-resonance formula and
numerical integration results using a resonance strength of 60 neV yield a
reaction rate approximately twice as large as the ``High'' rate of Ref. \cite%
{Ja2001} at the lower temperatures where their ``High'' rate is due mostly
to this resonance. The reason for this difference is unknown, but the $%
\delta $-resonance formula and numerical integration results were verified
by a third technique \cite{Bl2002} in which the reaction rate was calculated
using a Breit-Wigner resonance shape.

Another effect that has been overlooked in previous evaluations of the $%
^{22} $Ne($\alpha $,\textit{n})$^{25}$Mg reaction rate is the uncertainty
due to the resonance energy. As limits for resonance strengths are pushed
lower and lower, the uncertainty in the energy of the resonance can become a
significant effect. For example, in Ref. \cite{Ja2001} the energy of the
possible $E_{\alpha }$=635-keV resonance was estimated to be uncertain by $%
\pm $10 keV. Using Eq. \ref{ResReacRate}, this uncertainty in the resonance
energy translates to a factor of 2.7 uncertainty in the reaction rate at $%
T_{9}$=0.2, which is comparable to the total uncertainty of a factor of 5.9
(``High''/``Low'') at this temperature recommended in Ref. \cite{Ja2001}.
Although it is questionable that the energy of this state so uncertain, the
results of the present work should make it clear that energies resulting
from analysis of the neutron data are so precise that this source of
uncertainty now is practically eliminated.

\section{Summary and conclusions}

I have analyzed previously reported $^{\text{nat}}$Mg+\textit{n} total and $%
^{25}$Mg(\textit{n},$\gamma $) cross sections to obtain parameters for
resonances below $E_{n}$=500 keV. With a few notable exceptions, the
obtained $^{24}$Mg+\textit{n} and $^{26}$Mg+\textit{n} parameters are in
agreement with previous results to within the experimental uncertainties.
The main focus of the present work has been to obtain an improved set of $%
^{25}$Mg+\textit{n} parameters and hence an improved estimate of the
uncertainty in the $^{22}$Ne($\alpha $,\textit{n})$^{25}$Mg astrophysical
reaction rate. This reaction is the main neutron source during the weak
component of the \textit{s}-process nucleosynthesis as well as a secondary
neutron source during the main component of the \textit{s} process. The
uncertainty in this rate at \textit{s}-process temperatures is dominated by
possible contributions from resonances between threshold and the lowest
observed resonance.

The new $^{25}$Mg+\textit{n} parameter set represents a substantial
improvement over previous work. For example, several $J^{\pi }$ assignments
were made and the partial widths for most resonances were determined. In the
previous analysis, no definite $J^{\pi }$ assignments were made and very few
partial widths were reported. Also, one previously reported \cite{We76}
resonance was not observed and, if it does exist, has a width much smaller
than reported in compilations \cite{Mu81,En98}. In addition, four new
resonances were observed in this energy range. Furthermore, corresponding
resonances were found for all three of the $^{22}$Ne($\alpha $,\textit{n})$%
^{25}$Mg resonances as well as the four $^{26}$Mg($\gamma $,\textit{n})$%
^{25} $Mg resonances reported \cite{Ja2001,Be69} in this energy range,
although the energy or width of the lowest $^{22}$Ne($\alpha $,\textit{n})$%
^{25}$Mg resonances appears to be in error.

Only natural parity states in $^{26}$Mg can contribute to the $^{22}$Ne($%
\alpha $,\textit{n})$^{25}$Mg reaction rate. Much attention has been focused
on a $^{26}$Mg($\gamma $,\textit{n})$^{25}$Mg resonance at $E_{L}$=54.3 keV
because it very likely has natural parity and therefore could correspond to
a $^{22}$Ne($\alpha $,\textit{n})$^{25}$Mg resonance at $E_{\alpha }$=636
keV, nearly the optimal energy to make a large contribution to the reaction
rate at \textit{s}-process temperatures. The $\mathcal{R}$-matrix analysis
of the present work revealed that of the 16 states observed, there are at
least two, and very likely three, other definite natural parity states in $%
^{26}$Mg in this energy range and two definite non-natural parity states.
The parameters for these natural parity states, together with yield limits
from a recent $^{22}$Ne($\alpha $,\textit{n})$^{25}$Mg measurement \cite%
{Ja2001} have been used to estimate the contributions of these states to
this reaction rate. In a recent report \cite{Ja2001}, only one of these
states ($E_{\alpha }$=636 keV) was considered, and it was concluded that the
uncertainty in the reaction rate was much less than previously estimated.
However, using the upper limit on the resonance strength of the possible $%
E_{\alpha }$=636-keV resonance reported in this reference, I calculate that
they have underestimated the uncertainty due to this resonance alone by a
factor of two. More importantly, the definite natural parity resonance at $%
E_{n}$=19.880 keV, which corresponds to a possible $^{22}$Ne($\alpha $,%
\textit{n})$^{25}$Mg resonance at $E_{\alpha }$=588 keV, contributes a ten
times larger uncertainty to the rate at \textit{s}-process temperatures.

There are still at least 10 more states in $^{26}$Mg observed in the neutron
data that could contribute to the $^{22}$Ne($\alpha $,\textit{n})$^{25}$Mg
reaction rate. It was not possible to make definite $J^{\pi }$ assignments
for these resonances because $^{25}$Mg comprises only 10\% of the $^{nat}$Mg
sample used in the total cross section measurements analyzed in this work.
New high-resolution total cross section measurements on highly-enriched $%
^{25}$Mg samples could go a long way towards discerning how many of these
states have natural parity. It may be that neutron elastic scattering
measurements would also be needed in the more difficult cases. In addition,
it would be useful to determine the energy and width of the lowest observed $%
^{22}$Ne($\alpha $,\textit{n})$^{25}$Mg resonance with improved precision.
At present, the reported energy is not in good agreement with any observed
neutron resonance, and the reported width implies that a different state at
nearly the same energy has been observed in $^{22}$Ne($\alpha $,$\gamma $)$%
^{26}$Mg measurements.

\begin{acknowledgments}
I would like to thank J. A. Harvey, R. L. Macklin, and D. Wiarda for help
locating the data analyzed in this work, and D. W. Bardayan, J. C. Blackmon,
J. A. Harvey, and S. Raman for useful discussions. This research was
supported by the U.S. Department of Energy under Contract No.
DE-AC05-00OR22725 with UT-Battelle, LLC.
\end{acknowledgments}

\newif\ifabfull\abfulltrue


\begin{thebibliography}{30}
\expandafter\ifx\csname natexlab\endcsname\relax\def\natexlab#1{#1}\fi
\expandafter\ifx\csname bibnamefont\endcsname\relax
  \def\bibnamefont#1{#1}\fi
\expandafter\ifx\csname bibfnamefont\endcsname\relax
  \def\bibfnamefont#1{#1}\fi
\expandafter\ifx\csname citenamefont\endcsname\relax
  \def\citenamefont#1{#1}\fi
\expandafter\ifx\csname url\endcsname\relax
  \def\url#1{\texttt{#1}}\fi
\expandafter\ifx\csname urlprefix\endcsname\relax\def\urlprefix{URL }\fi
\providecommand{\bibinfo}[2]{#2}
\providecommand{\eprint}[2][]{\url{#2}}

\bibitem[{\citenamefont{Kappeler}(1999)}]{Ka99b}
\bibinfo{author}{\bibfnamefont{F.}~\bibnamefont{Kappeler}},
  \bibinfo{journal}{Prog. in Particle and Nucl. Phys.}
  \textbf{\bibinfo{volume}{43}}, \bibinfo{pages}{419} (\bibinfo{year}{1999}).

\bibitem[{\citenamefont{Hoffman et~al.}(2001)\citenamefont{Hoffman, Woosley,
  and Weaver}}]{Ho2001}
\bibinfo{author}{\bibfnamefont{R.~D.} \bibnamefont{Hoffman}},
  \bibinfo{author}{\bibfnamefont{S.~E.} \bibnamefont{Woosley}},
  \bibnamefont{and} \bibinfo{author}{\bibfnamefont{T.~A.}
  \bibnamefont{Weaver}}, \bibinfo{journal}{Astrophys. J.}
  \textbf{\bibinfo{volume}{549}}, \bibinfo{pages}{1085} (\bibinfo{year}{2001}).

\bibitem[{\citenamefont{Straniero et~al.}(1995)\citenamefont{Straniero,
  Gallino, Busso, Chieffi, Raiteri, Limongi, and Salaris}}]{St95}
\bibinfo{author}{\bibfnamefont{O.}~\bibnamefont{Straniero}},
  \bibinfo{author}{\bibfnamefont{R.}~\bibnamefont{Gallino}},
  \bibinfo{author}{\bibfnamefont{M.}~\bibnamefont{Busso}},
  \bibinfo{author}{\bibfnamefont{A.}~\bibnamefont{Chieffi}},
  \bibinfo{author}{\bibfnamefont{R.~M.} \bibnamefont{Raiteri}},
  \bibinfo{author}{\bibfnamefont{M.}~\bibnamefont{Limongi}}, \bibnamefont{and}
  \bibinfo{author}{\bibfnamefont{M.}~\bibnamefont{Salaris}},
  \bibinfo{journal}{Astrophys. J.} \textbf{\bibinfo{volume}{440}},
  \bibinfo{pages}{L85} (\bibinfo{year}{1995}).

\bibitem[{\citenamefont{Costa et~al.}(2000)\citenamefont{Costa, Rayet, Zappala,
  and Arnould}}]{Co2000}
\bibinfo{author}{\bibfnamefont{V.}~\bibnamefont{Costa}},
  \bibinfo{author}{\bibfnamefont{M.}~\bibnamefont{Rayet}},
  \bibinfo{author}{\bibfnamefont{R.~A.} \bibnamefont{Zappala}},
  \bibnamefont{and} \bibinfo{author}{\bibfnamefont{M.}~\bibnamefont{Arnould}},
  \bibinfo{journal}{Astron. and Astrophys.} \textbf{\bibinfo{volume}{358}},
  \bibinfo{pages}{L67} (\bibinfo{year}{2000}).

\bibitem[{\citenamefont{Drotleff et~al.}(1991)\citenamefont{Drotleff, Denker,
  Hammer, Knee, Kuchler, Streit, Rolfs, and Trautvetter}}]{Dr91}
\bibinfo{author}{\bibfnamefont{H.~W.} \bibnamefont{Drotleff}},
  \bibinfo{author}{\bibfnamefont{A.}~\bibnamefont{Denker}},
  \bibinfo{author}{\bibfnamefont{J.~W.} \bibnamefont{Hammer}},
  \bibinfo{author}{\bibfnamefont{H.}~\bibnamefont{Knee}},
  \bibinfo{author}{\bibfnamefont{S.}~\bibnamefont{Kuchler}},
  \bibinfo{author}{\bibfnamefont{D.}~\bibnamefont{Streit}},
  \bibinfo{author}{\bibfnamefont{C.}~\bibnamefont{Rolfs}}, \bibnamefont{and}
  \bibinfo{author}{\bibfnamefont{H.~P.} \bibnamefont{Trautvetter}},
  \bibinfo{journal}{Z. Phys. A} \textbf{\bibinfo{volume}{338}},
  \bibinfo{pages}{367} (\bibinfo{year}{1991}).

\bibitem[{\citenamefont{Harms et~al.}(1991)\citenamefont{Harms, Kratz, and
  Wiescher}}]{Ha91}
\bibinfo{author}{\bibfnamefont{V.}~\bibnamefont{Harms}},
  \bibinfo{author}{\bibfnamefont{K.-L.} \bibnamefont{Kratz}}, \bibnamefont{and}
  \bibinfo{author}{\bibfnamefont{M.}~\bibnamefont{Wiescher}},
  \bibinfo{journal}{Phys. Rev. C} \textbf{\bibinfo{volume}{43}},
  \bibinfo{pages}{2849} (\bibinfo{year}{1991}).

\bibitem[{\citenamefont{Drotleff et~al.}(1993)\citenamefont{Drotleff, Denker,
  Knee, Soine, Wolf, Hammer, Greife, Rolfs, and Trautvetter}}]{Dr93}
\bibinfo{author}{\bibfnamefont{H.~W.} \bibnamefont{Drotleff}},
  \bibinfo{author}{\bibfnamefont{A.}~\bibnamefont{Denker}},
  \bibinfo{author}{\bibfnamefont{H.}~\bibnamefont{Knee}},
  \bibinfo{author}{\bibfnamefont{M.}~\bibnamefont{Soine}},
  \bibinfo{author}{\bibfnamefont{G.}~\bibnamefont{Wolf}},
  \bibinfo{author}{\bibfnamefont{J.~W.} \bibnamefont{Hammer}},
  \bibinfo{author}{\bibfnamefont{U.}~\bibnamefont{Greife}},
  \bibinfo{author}{\bibfnamefont{C.}~\bibnamefont{Rolfs}}, \bibnamefont{and}
  \bibinfo{author}{\bibfnamefont{H.~P.} \bibnamefont{Trautvetter}},
  \bibinfo{journal}{Astrophys. J.} \textbf{\bibinfo{volume}{414}},
  \bibinfo{pages}{735} (\bibinfo{year}{1993}).

\bibitem[{\citenamefont{Giesen et~al.}(1993)\citenamefont{Giesen, Browne,
  Gorres, Graff, Iliadis, Trautvetter, Wiescher, Harms, Kratz, Pfeiffer
  et~al.}}]{Gi93}
\bibinfo{author}{\bibfnamefont{U.}~\bibnamefont{Giesen}},
  \bibinfo{author}{\bibfnamefont{C.~P.} \bibnamefont{Browne}},
  \bibinfo{author}{\bibfnamefont{J.}~\bibnamefont{Gorres}},
  \bibinfo{author}{\bibfnamefont{S.}~\bibnamefont{Graff}},
  \bibinfo{author}{\bibfnamefont{C.}~\bibnamefont{Iliadis}},
  \bibinfo{author}{\bibfnamefont{H.-P.} \bibnamefont{Trautvetter}},
  \bibinfo{author}{\bibfnamefont{M.}~\bibnamefont{Wiescher}},
  \bibinfo{author}{\bibfnamefont{W.}~\bibnamefont{Harms}},
  \bibinfo{author}{\bibfnamefont{K.~L.} \bibnamefont{Kratz}},
  \bibinfo{author}{\bibfnamefont{B.}~\bibnamefont{Pfeiffer}},
  \bibnamefont{et~al.}, \bibinfo{journal}{Nucl. Phys.}
  \textbf{\bibinfo{volume}{A561}}, \bibinfo{pages}{95} (\bibinfo{year}{1993}).

\bibitem[{\citenamefont{Jaeger et~al.}(2001)\citenamefont{Jaeger, Kunz, Mayer,
  Hammer, Staudt, Kratz, and Pfeiffer}}]{Ja2001}
\bibinfo{author}{\bibfnamefont{M.}~\bibnamefont{Jaeger}},
  \bibinfo{author}{\bibfnamefont{R.}~\bibnamefont{Kunz}},
  \bibinfo{author}{\bibfnamefont{A.}~\bibnamefont{Mayer}},
  \bibinfo{author}{\bibfnamefont{J.~W.} \bibnamefont{Hammer}},
  \bibinfo{author}{\bibfnamefont{G.}~\bibnamefont{Staudt}},
  \bibinfo{author}{\bibfnamefont{K.~L.} \bibnamefont{Kratz}}, \bibnamefont{and}
  \bibinfo{author}{\bibfnamefont{B.}~\bibnamefont{Pfeiffer}},
  \bibinfo{journal}{Phys. Rev. Lett.} \textbf{\bibinfo{volume}{87}},
  \bibinfo{pages}{202501} (\bibinfo{year}{2001}).

\bibitem[{\citenamefont{Berman et~al.}(1969)\citenamefont{Berman, Hemert, and
  Bowman}}]{Be69}
\bibinfo{author}{\bibfnamefont{B.~L.} \bibnamefont{Berman}},
  \bibinfo{author}{\bibfnamefont{R.~L.~V.} \bibnamefont{Hemert}},
  \bibnamefont{and} \bibinfo{author}{\bibfnamefont{C.~D.}
  \bibnamefont{Bowman}}, \bibinfo{journal}{Phys. Rev. Lett.}
  \textbf{\bibinfo{volume}{23}}, \bibinfo{pages}{386} (\bibinfo{year}{1969}).

\bibitem[{\citenamefont{Caughlan and Fowler}(1988)}]{Ca88}
\bibinfo{author}{\bibfnamefont{G.~R.} \bibnamefont{Caughlan}} \bibnamefont{and}
  \bibinfo{author}{\bibfnamefont{W.~A.} \bibnamefont{Fowler}},
  \bibinfo{journal}{Atomic Data Nucl. Data Tables}
  \textbf{\bibinfo{volume}{40}}, \bibinfo{pages}{282} (\bibinfo{year}{1988}).

\bibitem[{\citenamefont{Kappeler et~al.}(1994)\citenamefont{Kappeler, Wiescher,
  Giesen, Gorres, Baraffe, Eid, Raiteri, Busso, Gallino, Limongi
  et~al.}}]{Ka94}
\bibinfo{author}{\bibfnamefont{F.}~\bibnamefont{Kappeler}},
  \bibinfo{author}{\bibfnamefont{M.}~\bibnamefont{Wiescher}},
  \bibinfo{author}{\bibfnamefont{U.}~\bibnamefont{Giesen}},
  \bibinfo{author}{\bibfnamefont{J.}~\bibnamefont{Gorres}},
  \bibinfo{author}{\bibfnamefont{I.}~\bibnamefont{Baraffe}},
  \bibinfo{author}{\bibfnamefont{M.~E.} \bibnamefont{Eid}},
  \bibinfo{author}{\bibfnamefont{C.~M.} \bibnamefont{Raiteri}},
  \bibinfo{author}{\bibfnamefont{M.}~\bibnamefont{Busso}},
  \bibinfo{author}{\bibfnamefont{R.}~\bibnamefont{Gallino}},
  \bibinfo{author}{\bibfnamefont{M.}~\bibnamefont{Limongi}},
  \bibnamefont{et~al.}, \bibinfo{journal}{Astrophys. J.}
  \textbf{\bibinfo{volume}{437}}, \bibinfo{pages}{396} (\bibinfo{year}{1994}).

\bibitem[{\citenamefont{Angulo et~al.}(1999)\citenamefont{Angulo, Arnould,
  Rayet, Descouvemont, Baye, Leclercq-Willain, Coc, Barhoumi, Aguer, Rolfs
  et~al.}}]{An99}
\bibinfo{author}{\bibfnamefont{C.}~\bibnamefont{Angulo}},
  \bibinfo{author}{\bibfnamefont{M.}~\bibnamefont{Arnould}},
  \bibinfo{author}{\bibfnamefont{M.}~\bibnamefont{Rayet}},
  \bibinfo{author}{\bibfnamefont{P.}~\bibnamefont{Descouvemont}},
  \bibinfo{author}{\bibfnamefont{D.}~\bibnamefont{Baye}},
  \bibinfo{author}{\bibfnamefont{C.}~\bibnamefont{Leclercq-Willain}},
  \bibinfo{author}{\bibfnamefont{A.}~\bibnamefont{Coc}},
  \bibinfo{author}{\bibfnamefont{S.}~\bibnamefont{Barhoumi}},
  \bibinfo{author}{\bibfnamefont{P.}~\bibnamefont{Aguer}},
  \bibinfo{author}{\bibfnamefont{C.}~\bibnamefont{Rolfs}},
  \bibnamefont{et~al.}, \bibinfo{journal}{Nucl. Phys.}
  \textbf{\bibinfo{volume}{A656}}, \bibinfo{pages}{3} (\bibinfo{year}{1999}).

\bibitem[{\citenamefont{Mughabghab et~al.}(1981)\citenamefont{Mughabghab,
  Divadeenam, and Holden}}]{Mu81}
\bibinfo{author}{\bibfnamefont{S.~F.} \bibnamefont{Mughabghab}},
  \bibinfo{author}{\bibfnamefont{M.}~\bibnamefont{Divadeenam}},
  \bibnamefont{and} \bibinfo{author}{\bibfnamefont{N.~E.}
  \bibnamefont{Holden}}, \emph{\bibinfo{title}{Neutron Cross Sections}},
  vol.~\bibinfo{volume}{1} (\bibinfo{publisher}{Academic},
  \bibinfo{address}{New York}, \bibinfo{year}{1981}).

\bibitem[{\citenamefont{Endt}(1998)}]{En98}
\bibinfo{author}{\bibfnamefont{P.~M.} \bibnamefont{Endt}},
  \bibinfo{journal}{Nucl. Phys.} \textbf{\bibinfo{volume}{A633}},
  \bibinfo{pages}{1} (\bibinfo{year}{1998}).

\bibitem[{\citenamefont{Newson et~al.}(1959)\citenamefont{Newson, Block,
  Nichols, Taylor, and Furr}}]{Ne59}
\bibinfo{author}{\bibfnamefont{H.~W.} \bibnamefont{Newson}},
  \bibinfo{author}{\bibfnamefont{R.~C.} \bibnamefont{Block}},
  \bibinfo{author}{\bibfnamefont{P.~F.} \bibnamefont{Nichols}},
  \bibinfo{author}{\bibfnamefont{A.}~\bibnamefont{Taylor}}, \bibnamefont{and}
  \bibinfo{author}{\bibfnamefont{A.~K.} \bibnamefont{Furr}},
  \bibinfo{journal}{Annals of Physics} \textbf{\bibinfo{volume}{8}},
  \bibinfo{pages}{211} (\bibinfo{year}{1959}).

\bibitem[{\citenamefont{Singh et~al.}(1974)\citenamefont{Singh, Liou,
  Rainwater, and Hacken}}]{Si74}
\bibinfo{author}{\bibfnamefont{U.~N.} \bibnamefont{Singh}},
  \bibinfo{author}{\bibfnamefont{H.~I.} \bibnamefont{Liou}},
  \bibinfo{author}{\bibfnamefont{J.}~\bibnamefont{Rainwater}},
  \bibnamefont{and} \bibinfo{author}{\bibfnamefont{G.}~\bibnamefont{Hacken}},
  \bibinfo{journal}{Phys. Rev. C} \textbf{\bibinfo{volume}{10}},
  \bibinfo{pages}{2150} (\bibinfo{year}{1974}).

\bibitem[{\citenamefont{Weigmann et~al.}(1976)\citenamefont{Weigmann, Macklin,
  and Harvey}}]{We76}
\bibinfo{author}{\bibfnamefont{H.}~\bibnamefont{Weigmann}},
  \bibinfo{author}{\bibfnamefont{R.~L.} \bibnamefont{Macklin}},
  \bibnamefont{and} \bibinfo{author}{\bibfnamefont{J.~A.}
  \bibnamefont{Harvey}}, \bibinfo{journal}{Phys. Rev. C}
  \textbf{\bibinfo{volume}{14}}, \bibinfo{pages}{1328} (\bibinfo{year}{1976}).

\bibitem[{\citenamefont{Larson}(2000)}]{La2000}
\bibinfo{author}{\bibfnamefont{N.~M.} \bibnamefont{Larson}},
  \bibinfo{type}{Tech. Rep.} \bibinfo{number}{ORNL/TM-2000/252},
  \bibinfo{institution}{Oak Ridge National Laboratory, 2000}
  (\bibinfo{year}{2000}).

\bibitem[{\citenamefont{Peelle et~al.}(1982)\citenamefont{Peelle, Harvey,
  Maienschein, Weston, Olsen, Larson, and Macklin}}]{Pe82}
\bibinfo{author}{\bibfnamefont{R.~W.} \bibnamefont{Peelle}},
  \bibinfo{author}{\bibfnamefont{J.~A.} \bibnamefont{Harvey}},
  \bibinfo{author}{\bibfnamefont{F.~C.} \bibnamefont{Maienschein}},
  \bibinfo{author}{\bibfnamefont{L.~W.} \bibnamefont{Weston}},
  \bibinfo{author}{\bibfnamefont{D.~K.} \bibnamefont{Olsen}},
  \bibinfo{author}{\bibfnamefont{D.~C.} \bibnamefont{Larson}},
  \bibnamefont{and} \bibinfo{author}{\bibfnamefont{R.~L.}
  \bibnamefont{Macklin}}, \bibinfo{type}{Tech. Rep.}
  \bibinfo{number}{ORNL/TM-8225}, \bibinfo{institution}{Oak Ridge National
  Laboratory} (\bibinfo{year}{1982}).

\bibitem[{\citenamefont{Bockhoff et~al.}(1990)\citenamefont{Bockhoff, Carlson,
  Wasson, Harvey, and Larson}}]{Bo90}
\bibinfo{author}{\bibfnamefont{K.~H.} \bibnamefont{Bockhoff}},
  \bibinfo{author}{\bibfnamefont{A.~D.} \bibnamefont{Carlson}},
  \bibinfo{author}{\bibfnamefont{O.~A.} \bibnamefont{Wasson}},
  \bibinfo{author}{\bibfnamefont{J.~A.} \bibnamefont{Harvey}},
  \bibnamefont{and} \bibinfo{author}{\bibfnamefont{D.~C.}
  \bibnamefont{Larson}}, \bibinfo{journal}{Nucl. Sci. and Eng.}
  \textbf{\bibinfo{volume}{106}}, \bibinfo{pages}{192} (\bibinfo{year}{1990}).

\bibitem[{\citenamefont{Reffo et~al.}(1997)\citenamefont{Reffo, A.Ventura, and
  Grandi}}]{Gu97b}
\bibinfo{editor}{\bibfnamefont{G.}~\bibnamefont{Reffo}},
  \bibinfo{editor}{\bibnamefont{A.Ventura}}, \bibnamefont{and}
  \bibinfo{editor}{\bibfnamefont{C.}~\bibnamefont{Grandi}}, eds.,
  \emph{\bibinfo{title}{International Conference on Nuclear Data for Science
  and Technology}} (\bibinfo{publisher}{Societa Italiana di Fisica},
  \bibinfo{address}{Bologna}, \bibinfo{year}{1997}).

\bibitem[{\citenamefont{Harvey and Wiarda}(2001)}]{Ha2001}
\bibinfo{author}{\bibfnamefont{J.~A.} \bibnamefont{Harvey}} \bibnamefont{and}
  \bibinfo{author}{\bibfnamefont{D.}~\bibnamefont{Wiarda}}
  (\bibinfo{year}{2001}), \bibinfo{note}{private communication}.

\bibitem[{\citenamefont{Macklin}(2001)}]{Ma2001}
\bibinfo{author}{\bibfnamefont{R.~L.} \bibnamefont{Macklin}}
  (\bibinfo{year}{2001}), \bibinfo{note}{private communication}.

\bibitem[{\citenamefont{Macklin and Winters}(1981)}]{Ma81}
\bibinfo{author}{\bibfnamefont{R.~L.} \bibnamefont{Macklin}} \bibnamefont{and}
  \bibinfo{author}{\bibfnamefont{R.~R.} \bibnamefont{Winters}},
  \bibinfo{journal}{Nucl. Sci. Eng.} \textbf{\bibinfo{volume}{78}},
  \bibinfo{pages}{110} (\bibinfo{year}{1981}).

\bibitem[{\citenamefont{Bowman et~al.}(1967)\citenamefont{Bowman, Sidhu, and
  Berman}}]{Bo67}
\bibinfo{author}{\bibfnamefont{C.~D.} \bibnamefont{Bowman}},
  \bibinfo{author}{\bibfnamefont{G.~S.} \bibnamefont{Sidhu}}, \bibnamefont{and}
  \bibinfo{author}{\bibfnamefont{B.~L.} \bibnamefont{Berman}},
  \bibinfo{journal}{Phys. Rev.} \textbf{\bibinfo{volume}{163}},
  \bibinfo{pages}{951} (\bibinfo{year}{1967}).

\bibitem[{\citenamefont{Wolke et~al.}(1989)\citenamefont{Wolke, Harms, Becker,
  Hammer, Kratz, Rolfs, Schroder, Trautvetter, Wiescher, and Wohr}}]{Wo89}
\bibinfo{author}{\bibfnamefont{K.}~\bibnamefont{Wolke}},
  \bibinfo{author}{\bibfnamefont{V.}~\bibnamefont{Harms}},
  \bibinfo{author}{\bibfnamefont{H.~W.} \bibnamefont{Becker}},
  \bibinfo{author}{\bibfnamefont{J.~W.} \bibnamefont{Hammer}},
  \bibinfo{author}{\bibfnamefont{K.~L.} \bibnamefont{Kratz}},
  \bibinfo{author}{\bibfnamefont{C.}~\bibnamefont{Rolfs}},
  \bibinfo{author}{\bibfnamefont{U.}~\bibnamefont{Schroder}},
  \bibinfo{author}{\bibfnamefont{H.~P.} \bibnamefont{Trautvetter}},
  \bibinfo{author}{\bibfnamefont{M.}~\bibnamefont{Wiescher}}, \bibnamefont{and}
  \bibinfo{author}{\bibfnamefont{A.}~\bibnamefont{Wohr}}, \bibinfo{journal}{Z.
  Phys. A} \textbf{\bibinfo{volume}{334}}, \bibinfo{pages}{491}
  (\bibinfo{year}{1989}).

\bibitem[{\citenamefont{Rolfs and Rodney}(1988)}]{Ro88}
\bibinfo{author}{\bibfnamefont{C.~E.} \bibnamefont{Rolfs}} \bibnamefont{and}
  \bibinfo{author}{\bibfnamefont{W.~S.} \bibnamefont{Rodney}},
  \emph{\bibinfo{title}{Cauldrons in the Cosmos}}
  (\bibinfo{publisher}{University of Chicago Press},
  \bibinfo{address}{Chicago}, \bibinfo{year}{1988}).

\bibitem[{\citenamefont{Bahcall and Fowler}(1969)}]{Ba69}
\bibinfo{author}{\bibfnamefont{N.~A.} \bibnamefont{Bahcall}} \bibnamefont{and}
  \bibinfo{author}{\bibfnamefont{W.~A.} \bibnamefont{Fowler}},
  \bibinfo{journal}{Astrophys. J.} \textbf{\bibinfo{volume}{157}},
  \bibinfo{pages}{659} (\bibinfo{year}{1969}).

\bibitem[{\citenamefont{Blackmon}(2002)}]{Bl2002}
\bibinfo{author}{\bibfnamefont{J.~C.} \bibnamefont{Blackmon}}
  (\bibinfo{year}{2002}), \bibinfo{note}{private communication}.

\end{thebibliography}
\end{document}